\documentclass[a4paper,10pt]{article}
\usepackage{graphicx}
\usepackage{amsmath}
\usepackage{amssymb}
\usepackage{theorem}
\usepackage{multicol}

\theorembodyfont{\rmfamily}  
\DeclareMathAlphabet{\bm}{OML}{cmm}{b}{it}
\newcommand{\ket}[1]{| #1 \rangle} 
\newcommand{\bra}[1]{\langle #1 |} 

\newcommand{\rom}[1]{\mathrm{#1}}
\newcommand{\bellstate}{{\cal \psi}}

\title{Security of  quantum key distribution protocol with 
two-way classical communication assisted by one-time pad encryption}
\author{
  Shun Watanabe 
 \thanks{shun-wata@it.ss.titech.ac.jp}
  \and
  Ryutaroh Matsumoto \thanks{ryutaroh@it.ss.titech.ac.jp}
  \and
  Tomohiko Uyematsu \thanks{uematsu@it.ss.titech.ac.jp} \and \\
Department of Communications and Integrated Systems, \\
Tokyo Institute of Technology, \\
2-12-1, Oookayama, Meguro-ku,Tokyo, 152-8552, Japan \\
Fax: +81-3-5734-2905
} 

\date{August 29, 2006}

\begin{document}
\maketitle
{\abstract
In this paper, we consider a quantum key distribution protocol (QKD)
with two-way classical communication that is assisted 
by one-time pad encryption.
We propose a two-way preprocessing that uses one-time pad encryption by
previously shared secret key,
and the net key rate of the QKD with proposed preprocessing exceeds the
key rate of the QKD without it. 
The preprocessing is reduced to
the entanglement distillation protocol with two-way classical
communication and
previously shared  EPR pairs
(two-way breeding protocol), and the security of QKD with the
preprocessing is guaranteed in the same way as
Shor and  Preskill's arguments. \\
\textbf{keyword}: quantum key distribution, entanglement distillation
protocol, two-way classical communication, one-time pad encryption
}

\section{Introduction}

Quantum key distribution (QKD) provides a way
for two parties Alice and Bob to share an unconditional secure key
in the presence of an eavesdropper Eve.
Unlike conventional schemes of key
distribution that rely on unproven computational assumptions,
the security of QDK is guaranteed by the principles of quantum mechanics.
Since an unknown quantum state cannot be cloned perfectly,
any eavesdropping attempt by 
Eve will disturb the transmitted 
quantum states.
Thus, by estimating the error rate of the transmitted quantum states,
Alice and Bob can estimate an amount of eavesdropping.
Then, by procedures such as 
the error correction and the privacy amplification,
Alice and Bob distill the final secure key from the raw key whose 
partial information is known to Eve.
The best-known  QKDs are the Bennett-Brassard 1984 (BB84)
protocol \cite{bennett84} or the six-state protocol \cite{bruss98}.
The security of the BB84 protocol was proved in \cite{biham00, mayers01},
and a simple proof was shown by Shor and Preskill in \cite{shor00},
in which the security of the protocol is proved by relating the
protocol to the entanglement distillation protocol (EDP) 
\cite{bennett96a,bennett96b, lo99} via 
Calderbank-Shor-Stean (CSS) quantum error correcting code 
\cite{calderbank96, steane96}.
After that, the security of the six-state protocol
was proved in \cite{lo01}.

In addition to the security of QKD, it is important to
increase the key rate of the QKD, where
the key rate is defined by the ratio of the length of
the final secure key to the length of  the raw key.
In \cite{gottesman03}, a preprocessing with two-way classical communication
was proposed in order to increase the key rate or the tolerable error rate of
the QKD, where the tolerable error rate is the error rate at which
the key rate becomes zero.
The security of QKD with two-way preprocessing is proved by relating 
the protocol to the EDP with two-way classical communication.
By this preprocessing, the key rate of the QKD is increased when
the noise of the channel is rather high.
Indeed, the tolerable error rate of the BB84 protocol is increased
from $11$ \% to $18.9$ \%, and that of the six-state protocol is
increased from $12.7$ \% to $26.4$ \%.
Later, it was shown that the BB84 protocol can tolerate $20.0$ \% 
error rate and the six-state protocol can tolerate $27.6$ \% error rate
in \cite{chau02}.
Since the distillation rate of 
the known  two-way EDPs exceed that of
one-way EDPs
only when the fidelity between
an initial mixed state and the EPR pair is rather low 
\cite{bennett96b},
the two-way preprocessing in the QKD is effective only when
the error rate of the channel is rather high.

In \cite{vollbrecht05}, a new type of two-way EDP was proposed.
This protocol uses previously shared  EPR pairs as assistant resource, and
the distillation rate of this EDP exceeds that of one-way EDPs
for whole range of  the fidelity. 
Motivated by \cite{vollbrecht05}, 
we propose a two-way preprocessing for QKD that uses
one-time pad encryption by previously 
shared secret key.
The proposed preprocessing is related to the two-way EDP with 
previously shared 
EPR pairs, and the security of the QKD with proposed
preprocessing is guaranteed in the same way as 
\cite{gottesman03, shor00}.
The advantage of the proposed preprocessing is that
the net key rate of the QKD with proposed preprocessing
exceeds the key rate of one-way QKD
 even when the error rate of
the channel is rather low,
where the net key rate is defined by the key rate subtracted by
the ratio of the length of the consumed secret key in the protocol to
the length of raw key.
It should be noted that the use of one-time pad encryption
in the QKD is already proposed in the literature \cite{koashi03}
in order to simplify the analysis of the security.
In contrast to \cite{koashi03}, we introduced one-time pad encryption
in order to increase the net key rate of the QKD.

The rest of this paper is organized as follows.
In Section \ref{preliminaries},
we present the notations used throughout this paper (Section \ref{notation})
and review known QKD protocols (Section \ref{known-protocol}).
In Section \ref{preprocessing-with-secret-key}, we propose
general two-way preprocessing that uses one-time pad encryption,
and show the security of the QKD with proposed preprocessing.
In Section \ref{secure-key-rate-of-proposed-protocol},
we present a specific instance of proposed preprocessing,
and for six-state protocol
we compare the net key rate of the QKD with 
proposed preprocessing, the key rate of the QKD with only one-way
classical communication,
and the key rate of the QKD with conventional two-way preprocessing.

%%%%%%%%%
\section{Preliminaries}
\label{preliminaries}

\subsection{Notations}
\label{notation}

In this section, we present the notations
used throughout this paper.
We denote two-dimensional Hilbert space (qubit) by 
${\cal H}$. 
In this paper, we use three orthonormal bases of ${\cal H}$:
$\{ \ket{0}, \ket{1} \}$, 
$\{ \ket{\overline{0}} = \frac{1}{\sqrt{2}}(\ket{0}+\ket{1}),
\ket{\overline{1}} = \frac{1}{\sqrt{2}}(\ket{0}-\ket{1}) \}$,
and $\{ \ket{\overline{\overline{0}}} = \frac{1}{\sqrt{2}}(\ket{0}+i\ket{1}),
\ket{\overline{\overline{1}}} = \frac{1}{\sqrt{2}}(\ket{0}-i\ket{1}) \}$.
For a two-qubits Hilbert space 
${\cal H}^{\otimes 2} = {\cal H} \otimes {\cal H}$, there
exists four maximally entangled states called Bell states:
\begin{eqnarray*}
\ket{\bellstate_{00}} &=& \frac{1}{\sqrt{2}}(\ket{00}+\ket{11}), \\
\ket{\bellstate_{10}} &=& \frac{1}{\sqrt{2}}(\ket{01}+\ket{10}), \\
\ket{\bellstate_{01}} &=& \frac{1}{\sqrt{2}}(\ket{00}-\ket{11}), \\
\ket{\bellstate_{11}} &=& \frac{1}{\sqrt{2}}(\ket{01}-\ket{10}).
\end{eqnarray*}
The projectors onto these Bell states are denoted by
$\mathsf{P}_{ij} = \ket{\bellstate_{ij}}\bra{\bellstate_{ij}}$. 
For vectors $\bm{a} = (a_1,\ldots, a_n)$ and 
$\bm{b} = (b_1,\ldots,b_n)$, $\ket{\bellstate_{\bm{a}\bm{b}}^n}$ represents
\begin{eqnarray*}
\ket{\bellstate_{a_1 b_1}} \otimes \cdots \otimes \ket{\bellstate_{a_n b_n}}.
\end{eqnarray*}
The projector onto $\ket{\bellstate_{\bm{a}\bm{b}}^n}$ is denoted by
$\mathsf{P}_{\bm{a}\bm{b}}^n = 
\ket{\bellstate_{\bm{a}\bm{b}}^n}\bra{\bellstate_{\bm{a}\bm{b}}^n}$.

For a probability distribution $\{ p_i\}_{i=1}^m,~\sum_{i=1}^m p_i = 1$,
$H(p_1,\ldots,p_m)$ is the entropy function defined by
$H(p_1,\ldots,p_m) = \sum_{i=1}^m - p_i \log p_i$,
where the base of $\log$ is $2$.
For an $m$-tuple of non-negative numbers 
$\{ p_i\}_{i=1}^m$ with $\sum_{i=1}^m p_i = P$, 
$H[ p_1, \ldots, p_m ]$ denotes the entropy of the normalized
probability distribution, i.e.,
$H[ p_1,\ldots,p_m] = H(p_1/P,\ldots,p_m/P)$.
%%%%%
\subsection{Known protocols}
\label{known-protocol}

In this section, we review known protocols: the QKD with one-way
classical communication, and the QKD with the two-way preprocessing,
and the security of those protocols \cite{gottesman03, lo01, shor00}.
The prepare and measure
QKD protocols consist of two phases, the quantum transmission
phase and the key distillation phase.
In the quantum transmission phase, the sender Alice 
sends a random bit sequence by sending 
quantum states, with $\{ \ket{0}, \ket{1} \}$ basis or
$\{ \ket{\overline{0}},\ket{\overline{1}} \}$ basis 
in the BB84 protocol, and with $\{ \ket{0}, \ket{1} \}$ basis,
$\{ \ket{\overline{0}},\ket{\overline{1}} \}$ basis, or
$\{ \ket{\overline{\overline{0}}}, \ket{\overline{\overline{1}}} \}$
basis
in the six-state protocol.
Then, revealing part of shared bit sequences, 
Alice and Bob estimates error rates.
If estimated error rates are too high, then they abort the protocol.
In the end of this phase, Alice and Bob get raw keys respectively.
In the following, we consider the raw key $\bm{x}$ that is transmitted
by $\{ \ket{0}, \ket{1} \}$ basis, but the final secure key
can be distilled from the raw keys that are transmitted in other bases
in the same way.
The key distillation phase is further
divided into three part:
\begin{description}
\item[Two-way preprocessing]
Alice and Bob perform preprocessing in order to 
separate the raw key into two groups, one with 
higher bit error rate and one with lower bit error rate.
\item[Error correction]
Alice and Bob
eliminate the disagreement between Alice and Bob's raw keys
by an error-correcting code.
\item[Privacy amplification]
Alice and Bob
reduce the leaked information
about the raw key
to Eve by shortening the raw key into
a shorter bit sequence with a hash function.
\end{description} 
Finally, Alice and Bob  share a secret key 
$\bm{k}$.

The security of the final secret key $\bm{k}$ is shown as follows.
When Alice sends a randomly
chosen raw key $\bm{x} = (x_1,\ldots, x_n) \in \mathbb{F}_2^n$ 
by transmitting the quantum state
$\ket{\bm{x}} := \ket{x_1} \otimes \cdots \otimes \ket{x_n}$ to Bob, 
Bob receive a state $\rho_\bm{x}^B$ and 
Eve has a state 
$\rho_\bm{x}^E = \rom{Tr}_B \ket{\bm{x}_{BE}}\bra{\bm{x}_{BE}}$, where
$\ket{\bm{x}_{BE}}$ is a purification of $\rho_\bm{x}^B$ in 
Bob's system ${\cal H}_B = {\cal H}^{\otimes n}$ 
and Eve's system ${\cal H}_E$.
In Eve's point of view,
this situation can be regarded as follows
by using a quantum state
on Alice's system ${\cal H}_A = {\cal H}^{\otimes n}$,
Bob's system ${\cal H}_B$ and Eve's system ${\cal H}_E$:
\begin{eqnarray}
\label{state-raw-key}
\rho^{ABE} = \frac{1}{2^n} \sum_{\bm{x} \in \mathbb{F}_2^n}
\ket{\bm{x}}\bra{\bm{x}}_A \otimes \rho_\bm{x}^{BE},
\end{eqnarray} 
where $\rho_\bm{x}^{BE} = \ket{\bm{x}_{BE}}\bra{\bm{x}_{BE}}$.
Then, Eve has the system ${\cal H}_E$ of the state
$\rho^E = \rom{Tr}_{AB} \rho^{ABE}$.

Let tripartite state $\ket{\Psi_{ABE}}$ be
\begin{eqnarray}
\label{coherent-state-raw-key}
\ket{\Psi_{ABE}} = \frac{1}{\sqrt{2^n}}
\sum_{\bm{x} \in \mathbb{F}_2^n} \ket{\bm{x}}_A \otimes \ket{\bm{x}}_{BE}.
\end{eqnarray}
Even if we assume that the state in Alice, Bob, and Eve's
systems is $\ket{\Psi_{ABE}}$ instead of $\rho^{ABE}$,
there is no difference in Eve's point of view, since
$\rom{Tr}_{AB} \ket{\Psi_{ABE}}\bra{\Psi_{ABE}} = \rho^E$.
Furthermore, we can assume that the state 
$\sigma^{AB} = \rom{Tr}_E \ket{\Psi_{ABE}}\bra{\Psi_{ABE}}$ is 
diagonal in the Bell basis, i.e.,
\begin{eqnarray}
\sigma^{AB} = \sum_{\bm{a},\bm{b} \in \mathbb{F}_2^n} P_{\bm{a},\bm{b}} 
\ket{\bellstate_{\bm{a}\bm{b}}^n}\bra{\bellstate_{\bm{a}\bm{b}}^n},
\hspace{5mm}
\sum_{\bm{a},\bm{b} \in \mathbb{F}_2^n} P_{\bm{a},\bm{b}} = 1
\label{bell-diagonal}
\end{eqnarray}
by the following reason \cite{biham00, kraus05}.
If $\sigma^{AB}$ is not diagonal in the Bell basis,
then we can perform twirling \cite{bennett96b, hamada03}, 
and providing  ancilla systems for
twirling to Eve increases her information.

According to the security proof of \cite{gottesman03,shor00},
Eve's information about $\bm{k}$ is negligible if 
Alice and Bob can distill a bipartite state 
almost close to the perfect EPR pairs 
\begin{eqnarray*}
\ket{\bellstate_{00}^m} =
\frac{1}{\sqrt{2^m}} \sum_{k \in \mathbb{F}_2^m}
\ket{\bm{k}}_A \otimes \ket{\bm{k}}_B
\end{eqnarray*}
from the mixed bipartite state $\sigma^{AB}$
by the EDP corresponding to
the two-way preprocessing, the error correction, and
the privacy amplification.

%%%%%%%%%%%%%%%%%%
\section{Preprocessing with one-time pad encryption}
\label{preprocessing-with-secret-key}

In this section, we propose new preprocessing that uses
two-way classical communication and
one-time pad encryption  by previously shared  secret key.
Then, we show the security of the proposed preprocessing by
reducing the proposed preprocessing to the
two-way EDP with previously shared EPR pairs.

When Alice and Bob have raw keys $\bm{x}$ and $\tilde{\bm{x}}$ respectively,
our new preprocessing is executed as follows.
Alice calculates  
parities $\bm{x} M^T \in \mathbb{F}_2^l$ 
for a parity check matrix $M$ and sends it 
encrypted by previously shared secret key $\bm{s} \in \mathbb{F}_2^l$, i.e.,
Alice sends $\bm{x} M^T + \bm{s}$,
where $M^T$ denotes the transpose of the matrix $M$.
Then, Bob subtracts $\bm{s}$ from $\bm{x} M^T + \bm{s}$, and
calculates parities $\bm{t} = (\bm{x}-\tilde{\bm{x}}) M^T$ and
sends it to Alice without encryption.
The information $\bm{t} = (\bm{x}-\tilde{\bm{x}})M^T$ can be used in
the subsequent processings: the error correction and the 
privacy amplification.
The main difference between this preprocessing and the conventional
two-way preprocessing \cite{gottesman03} is that
the information about Alice's raw key is not revealed.

This preprocessing is reduced to the two-way EDP with previously shared
EPR pairs as follows.
In Eve's point of view, above situation can be regarded as follows by
using a quantum state on Alice's system ${\cal H}_A$,
Bob's system ${\cal H}_B$ and Eve's system ${\cal H}_E$. 
Before the preprocessing, the state is
of the form Eq.~(\ref{state-raw-key}).
In the preprocessing, 
Bob will obtain a parity $\bm{t}$ with probability 
\begin{eqnarray*}
P_{\bm{t}|\bm{x}} = \rom{Tr} \left[ (\Pi_{\bm{x},\bm{t}} 
\otimes I_E) \rho_\bm{x}^{BE}
(\Pi_{\bm{x},\bm{t}} \otimes I_E ) \right],
\end{eqnarray*}
where $\Pi_{\bm{x},\bm{t}}$ is a projection operator defined by
\begin{eqnarray*}
\Pi_{\bm{x},\bm{t}} = \sum_{\scriptstyle \bm{u} \in \mathbb{F}_2^n \atop
\scriptstyle \bm{u} M^T = \bm{t}}
\ket{\bm{x}+\bm{u}}\bra{\bm{x}+\bm{u}},
\end{eqnarray*}
and $I_E$ is the identity operator on ${\cal H}_E$.
Since we assumed $\sigma^{AB}$ is of the form Eq.~(\ref{bell-diagonal}),
$P_{\bm{t}|\bm{x}}$ does not depends on $\bm{x}$,
thus we denote $P_{\bm{t}|\bm{x}}$ by $P_\bm{t}$.
Since the parities of the difference
of Alice and Bob's raw key, $\bm{t} = (\bm{x}-\tilde{\bm{x}}) M^T$, 
is revealed to Eve, in Eve's point of view the state of 
Eq.~(\ref{state-raw-key}) becomes 
\begin{eqnarray}
\label{state-preprocessed-key}
\hat{\rho}_\bm{t}^{ABE} = \frac{1}{2^n} \sum_{\bm{x} \in \mathbb{F}_2^n}
\ket{\bm{x}}\bra{\bm{x}}_A \otimes \hat{\rho}_{\bm{x},\bm{t}}^{BE},
\end{eqnarray}
where 
\begin{eqnarray*}
\rho_{\bm{x},\bm{t}}^{BE} = 
\frac{1}{P_\bm{t}} (\Pi_{\bm{x},\bm{t}} \otimes I_E) \rho_x^{BE}
(\Pi_{\bm{x},\bm{t}} \otimes I_E ).
\end{eqnarray*}
Then, Eve has the system ${\cal H}_E$ of the
state 
$\hat{\rho}_\bm{t}^E = \rom{Tr}_{AB} \hat{\rho}_\bm{t}^{ABE} = \sum_{\bm{x} \in \mathbb{F}_2^n} \frac{1}{2^n} \hat{\rho}^E_{\bm{x},\bm{t}}$,
where $\hat{\rho}_{\bm{x},\bm{t}}^E = \rom{Tr}_B \hat{\rho}_{\bm{x},\bm{t}}^{BE}$.

This preprocessing is equivalent to the following two-way EDP that uses 
previously shared  EPR pairs as ancilla.
Alice and Bob start from the state of the form
Eq.~(\ref{coherent-state-raw-key}).
Alice and Bob perform parity check by CNOT operation with
$\sigma^{AB}$ as  source qubits and 
ancilla EPR pairs $\ket{\bellstate_{00}^l}$ as  target qubits.
Specifically, if the $(i,j)$ element $M_{ij}$ of $M$ is
$1$, then Alice and Bob each perform CNOT operation 
with $j$-th qubit pair of $\sigma^{AB}$ as  source qubits
and $i$-th ancilla EPR pair as  target qubits. 
After performing CNOT parity check, Alice and Bob 
measure the ancilla EPR pairs with $\{ \ket{0}, \ket{1} \}$ 
basis and get a measurement results $\bm{a}, \bm{b} \in \mathbb{F}_2^l$
respectively. Then, they compare $\bm{a}$ and $\bm{b}$ by two-way classical
communication and get the difference of the parity $\bm{t}$.
Here, the state of Eq.~(\ref{coherent-state-raw-key})
becomes, ignoring the normalization,
\begin{eqnarray*}
\ket{\widetilde{\Psi}_{ABE}} = 
(\Pi_\bm{t} \otimes I_E)
\ket{\Psi_{ABE}},
\end{eqnarray*}
where $\Pi_\bm{t}$ is a projection operator defined by
\begin{eqnarray*}
\Pi_\bm{t} = \sum_{\scriptstyle \bm{u},\bm{v} \in \mathbb{F}_2^n \atop 
\scriptstyle \bm{u} M^T = \bm{t}}
\mathsf{P}_{\bm{u}\bm{v}}^n,
\end{eqnarray*}
which is the projector onto the Bell states that
causes parities $\bm{t}$.
From the relation
\begin{eqnarray*}
\mathsf{P}_{00} + \mathsf{P}_{01} &=& \ket{00}\bra{00} +
\ket{11}\bra{11}, \\
\mathsf{P}_{10} + \mathsf{P}_{11} &=& \ket{01}\bra{01} +
\ket{10}\bra{10},
\end{eqnarray*}
we have
\begin{eqnarray*}
\lefteqn{ \Pi_\bm{t} = 
\sum_{\scriptstyle \bm{u},\bm{v} \in \mathbb{F}_2^n \atop 
\scriptstyle \bm{u} M^T = \bm{t}} \mathsf{P}_{\bm{u}\bm{v}}^n } \\
&=& \sum_{\scriptstyle \bm{u} \in \mathbb{F}_2^n \atop 
\scriptstyle \bm{u} M^T = \bm{t}}
\sum_{\bm{v} \in \mathbb{F}_2^n} \mathsf{P}_{\bm{u}\bm{v}}^n \\
&=& \sum_{\scriptstyle \bm{u} \in \mathbb{F}_2^n \atop 
\scriptstyle \bm{u} M^T = \bm{t}}
\sum_{\bm{y} \in \mathbb{F}_2^n} \ket{\bm{y}}\bra{\bm{y}} \otimes 
\ket{\bm{y}+\bm{u}}\bra{\bm{y}+\bm{u}} \\ 
&=& \sum_{ \bm{y} \in \mathbb{F}_2^n}
\ket{\bm{y}}\bra{\bm{y}} \otimes \Pi_{\bm{y},\bm{t}}.
\end{eqnarray*}
Thus, we have
\begin{eqnarray*}
\lefteqn{ \rom{Tr} \ket{\widetilde{\Psi}_{ABE}}\bra{\widetilde{\Psi}_{ABE}}
= \rom{Tr} (\Pi_\bm{t} \otimes I_E)
\ket{\Psi_{ABE}}\bra{\Psi_{ABE}} (\Pi_\bm{t} \otimes I_E)} \\
&=& \rom{Tr} 
\left( \sum_{\bm{y} \in \mathbb{F}_2^n} 
\ket{\bm{y}}\bra{\bm{y}} \otimes \Pi_{\bm{y},\bm{t}} \otimes
I_E \right) \ket{\Psi_{ABE}}\bra{\Psi_{ABE}} 
\left( \sum_{\bm{y}^\prime \in \mathbb{F}_2^n} 
\ket{\bm{y}^\prime}\bra{\bm{y}^\prime} \otimes 
\Pi_{\bm{y}^\prime,\bm{t}} \otimes
I_E \right) \\
&=& \rom{Tr} \frac{1}{2^n}
\sum_{\bm{y},\bm{y}^\prime \in \mathbb{F}_2^n}
\ket{\bm{y}}\bra{\bm{y}^\prime} \otimes
(\Pi_{\bm{y},\bm{t}} \otimes I_E) \ket{\bm{y}_{BE}}\bra{\bm{y}^\prime_{BE}}
(\Pi_{\bm{y}^\prime,\bm{t}} \otimes I_E) \\
&=& \frac{1}{2^n} \sum_{\bm{y} \in \mathbb{F}_2^n}
\rom{Tr} (\Pi_{\bm{y},\bm{t}} \otimes I_E) 
\rho_\bm{y}^{BE} (\Pi_{\bm{y},\bm{t}} \otimes I_E) \\
&=& P_\bm{t}.
\end{eqnarray*}
Thus, $\ket{\widetilde{\Psi}_{ABE}}$ is normalized to
\begin{eqnarray*}
\ket{\widehat{\Psi}_{ABE}} = \frac{1}{\sqrt{P_\bm{t}}}
(\Pi_\bm{t} \otimes I_E) \ket{\Psi_{ABE}}.
\end{eqnarray*}
Then, Eve has the system ${\cal H}_E$ of the state
\begin{eqnarray*}
\lefteqn{ \rom{Tr}_{AB} \ket{\widehat{\Psi}_{ABE}}
\bra{\widehat{\Psi}_{ABE}} } \\
&=& \rom{Tr}_B \frac{1}{2^n} \sum_{\bm{x} \in \mathbb{F}_2^n}
\frac{1}{P_\bm{t}} (\Pi_{\bm{x},\bm{t}} \otimes I_E) \rho_\bm{x}^{BE}
(\Pi_{\bm{x},\bm{t}} \otimes I_E)  \\
&=& \hat{\rho}^E_\bm{t},
\end{eqnarray*}
which is same as the state in the QKD.
Consequently, combining with the argument in Section \ref{known-protocol},
a secret key that is derived by
the proposed preprocessing followed by the two-way preprocessing,
the error correction, and the privacy amplification is secure, if
Alice and Bob can distill a state almost close to the perfect EPR pairs
by the EDP with previously shared EPR pairs followed by the EDP
corresponding to the two-way preprocessing, the error correction,
and the privacy amplification.
%%%%%%%%%%%%%%%%%%%
\section{Six-state protocol with proposed preprocessing}
\label{secure-key-rate-of-proposed-protocol}

In this section, we present a specific instance of the preprocessing proposed
in Section \ref{preprocessing-with-secret-key}, 
and calculate the net key rate
of the six-state protocol
with proposed preprocessing,
where the net key rate is the ratio of the net key length to
the length of the raw key, and the net key length is
the difference between the length of the final secure key and 
the length of the secret key consumed in the
preprocessing. 
Our  protocol is executed as follows.
\begin{enumerate}
\renewcommand{\labelenumi}{(\theenumi)}
\item \label{six-state-step1}
Alice prepares $N$ qubits randomly chosen from
$\ket{0}$, $\ket{1}$, $\ket{\overline{0}}$,
$\ket{\overline{1}}$, $\ket{\overline{\overline{0}}}$,
and $\ket{\overline{\overline{1}}}$ and
sends them to Bob.
Bob acknowledges the receipt of the qubits and
measures them randomly along one of the following three bases:
$\{ \ket{0},\ket{1} \}$, 
$\{ \ket{\overline{0}}, \ket{\overline{1}} \}$, and
$\{ \ket{\overline{\overline{0}}}, \ket{\overline{\overline{1}}} \}$.
Using the  correspondence that
$\ket{0}$, $\ket{\overline{0}}$, and $\ket{\overline{\overline{0}}}$
represent $0$ while $\ket{1}$, $\ket{\overline{1}}$,
and $\ket{\overline{\overline{1}}}$ represent $1$,
Alice and Bob convert their preparation and 
measurement results  into binary sequence.
Then, Alice and Bob announce the bases they used to prepare and
measure each qubit.
They keep only those bits that are prepared and measured in the same basis.
\item Alice and Bob 
 divide their remaining binary sequence into three sets according to their
basis of measurement.
They randomly pick test bits from each set and publicly compare the
preparation and measurement results. Then they get the ratios of
errors $p_Z$, $p_X$, and $p_Y$ in each test bits.
If these error rates are too high to distill the secure key,
then they abort the protocol.
They calculate $q_X$, $q_Z$, and $q_Y$ from the relations
$p_Z = q_X + q_Y$, $p_X = q_Z + q_Y$, and 
$p_Y = q_X + q_Z$. 
The ratios of $X$ errors, $Z$ errors, and $Y$ errors in
untested qubits are close to $q_X$, $q_Z$, and $q_Y$ with
high probability.
If Alice and Bob perform a processing that are secure
for any uncorrelated Pauli channels with error rates close
to $(q_X, q_Y, q_Z)$, then 
the security of the final key distilled by the processing is
guaranteed \cite[Lemma 3]{gottesman03}.
\label{step-estimation}  
\item \label{new-preprocessing}
They distill the final secure keys from 
raw keys, that is, the binary sequences that are not
revealed in step (\ref{step-estimation}).
In the following, we consider the raw key $\bm{x}$ that is 
transmitted by $\{ \ket{0}, \ket{1} \}$ basis, but
we can distill secure keys from raw keys that are transmitted by
other bases in the same way.
Alice converts the raw key $\bm{x} = (x_1,\ldots,x_n)$
into $\bm{c}=(c_1,\ldots,c_{n/2})$, where 
$c_i = x_{2i-1} \oplus x_{2i}$.
Then, she calculates  parities $\bm{c} M^T \in \mathbb{F}_2^l$ for 
$\frac{n}{2} \times l$ parity check
matrix $M$.
Then she sends $\bm{c}M^T + \bm{s} \in \mathbb{F}_2^l$ to Bob,
where $\bm{s}$ is a previously shared secret key.
Similarly, Bob converts the raw key 
$\tilde{\bm{x}} =(\tilde{x}_1,\ldots,\tilde{x}_n)$ into
$\tilde{\bm{c}} = (\tilde{c}_1,\ldots, \tilde{c}_{n/2})$,
where $\tilde{c}_i = \tilde{x}_{2i-1} \oplus \tilde{x}_{2i}$.
Then he calculates parities $\bm{c} M^T - \tilde{\bm{c}}M^T$ and
sends it to Alice.
Since $\bm{c} - \tilde{\bm{c}}$ can be regarded as a binary error sequence
with error rate $P_{odd} = 2p_Z (1- p_Z)$,
Alice and Bob can identify $\bm{c} - \tilde{\bm{c}}$ from
$l \simeq \frac{n}{2}H(P_{even},P_{odd})$ 
parities \cite{bennett96a,vollbrecht05},
where $P_{even} = 1 - P_{odd}$.
Thus, Alice and Bob can know which blocks of length $2$ of 
$\bm{x} - \tilde{\bm{x}}$ has an even parity ($00$ or $11$), 
or an odd parity ($01$ or $10$).
\item \label{one-way}
Let $\bm{x}_e$ and $\tilde{\bm{x}}_e$ be sequences that consist of
the blocks with even parities.
For $\bm{x}_e$ and $\tilde{\bm{x}}_e$, Alice and Bob perform the
error correction with a linear code $C_1$ and 
the privacy amplification with a linear code $C_2$,
and get the final key in the same way as \cite{lo01, shor00}.
\item \label{two-way}
For each blocks with odd parities,
Alice and Bob announce the first bit of each blocks.
Then, Alice and Bob can identify the errors in the second bit
of each block, and Bob can correct them.
Let $\bm{x}_{0}$ and $\hat{\bm{x}}_{0}$ be sequences that consist of
second bit of blocks with 
$x_{2i-1} \oplus \tilde{x}_{2i-1} = 0$,
and $\bm{x}_{1}$ and $\hat{\bm{x}}_{1}$ be sequences that consist of
second bit of blocks with
$x_{2i-1} \oplus \tilde{x}_{2i-1} = 1$.
\item \label{one-way-2}
Since there is no more errors in $\hat{\bm{x}}_{0}$ and
$\hat{\bm{x}}_{1}$,  Alice and Bob can distill final secret keys
only by the privacy amplification.
\end{enumerate}

\begin{figure}[h]
    \includegraphics[width=\linewidth]{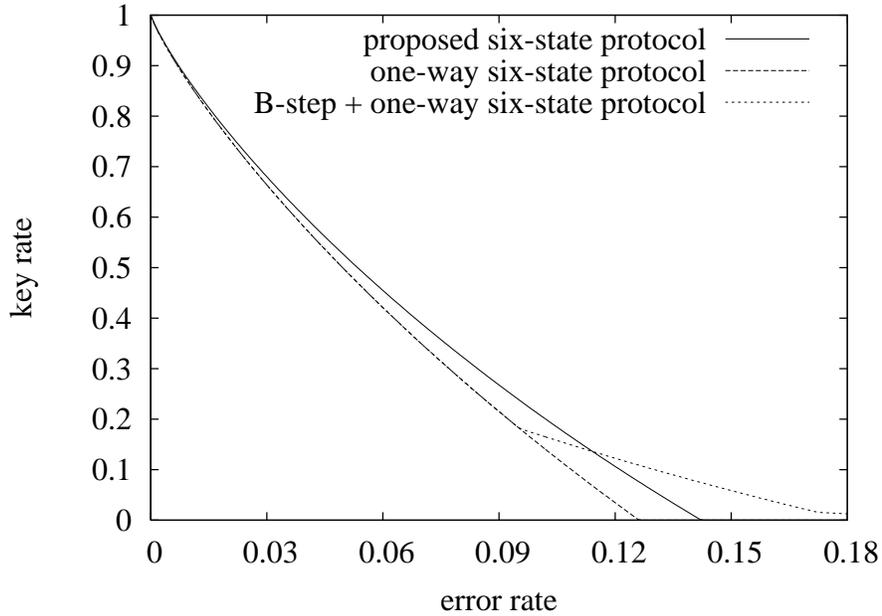}
\caption{Comparison among
the net key rate of the proposed protocol, the key rate of the six-state 
protocol with one-way classical communication \cite{lo01},
and the key rate of the six-state protocol with optimal number of
B-steps \cite{gottesman03}.}
\label{graph1}
\end{figure}

Each step of above protocol has the following meaning.
Step (\ref{new-preprocessing}) 
is two-way preprocessing with one-time pad encryption.
Step (\ref{two-way}) belongs to the class of 
conventional two-way preprocessing \cite{gottesman03},
and this step is reduced to the EDP in which Alice and Bob
measure $Z \otimes I$ for each blocks of qubits respectively.
Steps (\ref{one-way}) and 
(\ref{one-way-2}) are conventional error correction and
privacy amplification. 
The equivalent EDP of this protocol is the EDP proposed in
\cite{vollbrecht05}.
The net key rate of this protocol is exactly
the same as the distillation rate of the EDP in 
\cite{vollbrecht05}, which is calculated as follows.
The length of the secret key consumed in step  (\ref{new-preprocessing})
is  $\frac{n}{2} H(P_{even},P_{odd})$.
The length of the key that is distilled in step (\ref{one-way}) 
is $\frac{nP_{even}}{2}(2 - H[q_I^2,q_I q_Z, q_Z q_I, q_Z^2,q_X^2,q_X q_Y,
q_Y q_X, q_Y^2])$,
where $q_I = 1 - q_X - q_Y - q_Z$.
The length of the key that is distilled from $x_{0}$ and $x_{1}$
in step (\ref{one-way-2}) is
$\frac{n P_{odd}}{4}(1 - H[q_X, q_Y])$ and
$\frac{n P_{odd}}{4}(1 - H[q_I, q_Z])$ respectively.
Thus the net key rate is
\begin{eqnarray}
\lefteqn{ \frac{P_{even}}{2}(2 - 
H[q_I^2,q_I q_Z, q_Z q_I, q_Z^2,q_X^2,q_X q_Y, q_Y q_X, q_Y^2]) } \nonumber \\
&+&  \frac{ P_{odd}}{4}(1 - H[q_X, q_Y]) + 
\frac{ P_{odd}}{4}(1 - H[q_I, q_Z])
- \frac{1}{2} H(P_{even},P_{odd}) \nonumber \\
&=& 1 - H(q_I, q_X, q_Z, q_Y) +
\frac{P_{odd}}{4}\left\{
H[q_I, q_Z] + H[q_X, q_Y] \right\}.
\label{net-key-rate}
\end{eqnarray}
The net key rate of Eq.~(\ref{net-key-rate})
exceeds the key rate $1 - H(q_I, q_X, q_Z, q_Y)$
of the six-state protocol
with one-way classical communication \cite{lo01}.
The net key rate of proposed protocol and the key rate of
the six-state protocol with one-way classical communication are
compared in Fig.~\ref{graph1},
where we assumed the channel is the depolarizing channel with 
$q_X = q_Z = q_Y = p$, which indicates that
the estimated error rates are $p_Z = p_X = p_Y = 2p$.
The key rate of the six-state protocol that uses B-step of 
\cite[Section 7]{gottesman03}
optimal times before the error correction and the privacy amplification is also
plotted in Fig.~\ref{graph1}.
When the error rate of the channel is low,
the net key rate of 
the proposed protocol exceeds the key rate of the six-state protocol
with optimal number of B-steps. 

%%%%%%%%%%%%%%%%%
\section{BB84 protocol with proposed preprocessing}

In this section, we calculate the net key
rate of the BB84 protocol with proposed processing. 
The protocol is executed in almost the same way as the protocol
in Section \ref{secure-key-rate-of-proposed-protocol}
except the following two changes.

\begin{itemize}
\item Alice uses only $\ket{0}$, $\ket{1}$, $\ket{\overline{0}}$,
and $\ket{\overline{1}}$ in Step (\ref{six-state-step1}).

\item Alice and Bob can get ratios of errors $p_Z$ and $p_X$ in
Step (\ref{step-estimation}).
Thus, Alice and Bob have to perform a processing that is 
secure against the worst case of 
$q_X = p_Z - \alpha$, $q_Y = \alpha$, $q_Z = p_X - \alpha$
for a parameter $\alpha \in [0, \min\{ p_Z, p_X \}]$.
\end{itemize}

\begin{figure}[h]
    \includegraphics[width=\linewidth]{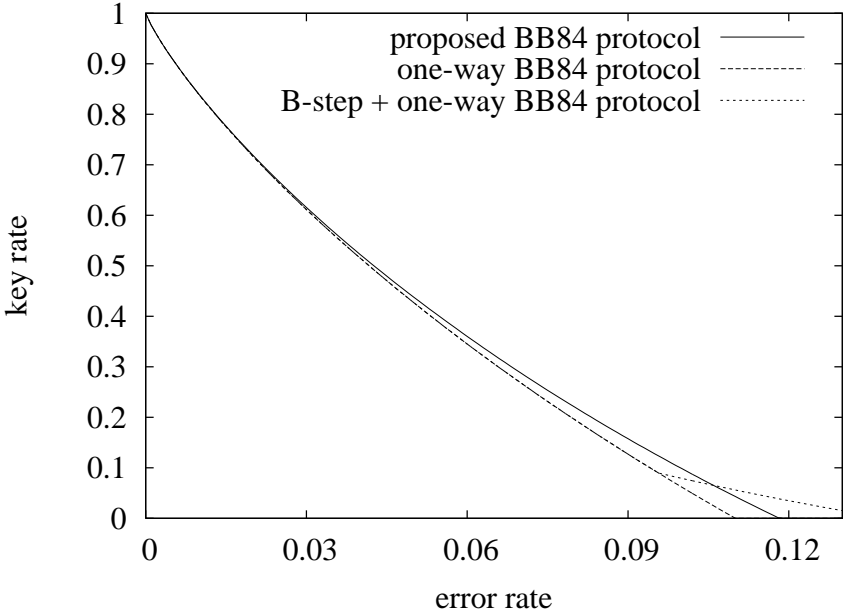}
\caption{Comparison among
the net key rate of the proposed protocol, the key rate of the BB84 
protocol with one-way classical communication,
and the key rate of the BB84 protocol with optimal number of
B-steps \cite{gottesman03}.}
\label{graph2}
\end{figure}

The net key rate of the BB84 protocol with proposed processing is
given by
\begin{eqnarray}
\min_{\alpha} \left[
 1 - H(q_I, q_X, q_Z, q_Y) +
\frac{P_{odd}}{4}\left\{
H[q_I, q_Z] + H[q_X, q_Y] \right\} \right].
\label{bb84-net-key-rate}
\end{eqnarray}
The net key rate of Eq.~(\ref{bb84-net-key-rate}) exceed
the key rate 
\begin{eqnarray*}
1 - h(p_Z) - h(p_X) = \min_{\alpha} \left[
1 - H(q_I, q_X, q_Z, q_Y) \right]
\end{eqnarray*}
of the BB84 protocol with one-way classical communication,
where $h(\cdot)$ is the binary entropy function.
The net key rate of proposed protocol and the key rate of the BB84 protocol
with one-way classical communication are compared in Fig.~\ref{graph2},
where we assumed $p_Z = p_X$.
The key rate of the BB84 protocol that uses B-step of 
\cite[Section 7]{gottesman03} optimal times before the error correction
and the privacy amplification is also plotted in Fig.~\ref{graph2}.
When the error rate of the channel is low,
the net key rate of 
the proposed protocol exceeds the key rate of the BB84 protocol
with optimal number of B-steps. 

%%%%%%%%%%%%
%%%%%%%%%%%%%%%%%%
\section{Conclusion}

In this paper, we proposed a two-way preprocessing that is 
assisted by one-time pad encryption, and showed that
proposed preprocessing is reduced to the
two-way EDP assisted by previously shared  EPR pairs,
and the security of the QKD with proposed preprocessing is guaranteed
in the same way as Shor and Preskill's arguments.
We also showed that for the six-state protocol and the BB84 protocol 
the net key rate of the QKD with proposed preprocessing 
exceeds the key rate of the QKD with only one-way classical communication,
and  also exceeds the key rate of 
the QKD with conventional two-way preprocessing 
when the error rate is low.

%%%%%%%%%%%%%%%%%
\section*{Acknowledgement}

This research is partly supported by the Japan Society
for the Promotion of Science under Grants-in-Aid for
Young Scientists No.~18760266.
%%%%%%%%%%%%%%%%%%%

\end{document}